\documentclass[12pt]{article}
\usepackage{amssymb,amsmath,epsfig}
\allowdisplaybreaks

\begin{document}
\title{\bf Greybody Factor for Quintessential Kerr-Newman Black Hole}
\author{M. Sharif \thanks{msharif.math@pu.edu.pk} and Qanitah Ama-Tul-Mughani
\thanks{qanitah94@gmail.com}\\
Department of Mathematics, University of the Punjab,\\
Quaid-e-Azam Campus, Lahore-54590, Pakistan.}

\date{}
\maketitle
\begin{abstract}
In this paper, we formulate an analytic expression of the greybody
factor for the Kerr-Newman black hole in the presence of the
quintessential field. Primarily, we analyze the profile of effective
potential by transforming the radial equation of motion into
standard Schr$\ddot{o}$dinger form through tortoise coordinate. The
two asymptotic solutions, in the form of hypergeometric functions,
are computed at distinct radial regions such as a black hole and
cosmological horizons determined by the quintessence. To extend the
viability over the whole radial regime, we match the analytical
solutions smoothly in an intermediate region by using a
semi-classical approach. We also calculate the emission rates and
absorption cross-section for the massless scalar fields to elaborate
on the significance of our result. It is found that the
electromagnetic force together with the gravitational pull of black
hole maximizes the effective potential and consequently, decreases
the emission process of scalar field particles.
\end{abstract}

\section{Introduction}

After the remarkable discovery of Hawking radiation \cite{1}, the
scattering of the material field from various black holes (BHs) has
become one of the interesting topics of strong gravitational fields.
These thermal spectra of radiations as the dividing line between
classical general relativity and quantum field theory may play an
effective role to resolve the mysterious nature of BHs. Due to
quantum mechanical effects, virtual particles created at the event
horizon spread out in the surrounding space that lead to decrease in
the BH mass and finally to its eventual evaporation. The geometry
outside the BH horizon has a significant impact on the emission rate
of scalar particles. In fact, the spacetime outside the boundary of
BH will work as a potential barrier for the emitted Hawking
radiation. Consequently, the radiation spectrum at the horizon is
exactly equal to the blackbody while the spectrum recorded by the
distant observer will depict different scenario. Mathematically, the
relative relation between the blackbody radiation and asymptotic
radiation spectra can be expressed as
\begin{equation}\nonumber
\gamma(w)=\left(\frac{|\tilde{\emph{A}}_{l,m}|^{2}
d^{3}k}{(e^{\frac{w}{T_{H}}}\pm1)(2\pi)^3}\right),
\end{equation}
where $T_{H}$ denotes the Hawking temperature and
$|\tilde{\emph{A}}_{l,m}|^{2}$ is dubbed as greybody factor which is
a frequency-dependent quantity. The greybody factor (rate of
absorption probability) is defined as the probability for an
incoming wave from infinity to be absorbed by the BH which is
directly related to the absorption cross-section \cite
{15'}-\cite{18}.

In astrophysics, the study of reflection as well as transmission
coefficients of the waves has gained much attention. Konoplya
\cite{a} computed the effective potential as well as quasinormal
modes associated with the decay of the massless scalar field for
small Schwarzschild-anti de Sitter BH. Konoplya and Zhidenko
\cite{b,c} analyzed the quasinormal modes of various BHs through
numerical and analytical techniques such as WKB method, integration
of the wavelike equations, Frobenius method, Fit and interpolation
approaches, etc. Ngampitipan and Boonserm \cite{41a} obtained
rigorous bounds on the transmission coefficient for
Reissner-Nordstr$\ddot{o}$n (RN) BH by using $2\times 2$ transfer
matrices. Boonserm et al. \cite{41'} established some bounds on the
absorption probability associated with scalar field excitations for
the Kerr-Newman BH. Toshmatov et al. \cite{42} computed the greybody
factors for regular BH spacetimes and found that charge parameter
decreases the transmission rate of the incident wave. Ahmed and
Saifullah \cite{19} studied the propagation of massless scalar
fields in the background of charged string theory and obtained an
analytic expression of the greybody factor in the low-energy
approximation. They also derived a general expression of absorption
probability for RN-de Sitter BH \cite{19'}. Recently, Dey and
Chakrabarti \cite{20} calculated quasinormal modes as well as
greybody factor for the Bardeen-de Sitter BH through electromagnetic
perturbations.

Recent astronomical observations  suggest that our universe is
expanding at an accelerating rate, driven by some unknown exotic
component dubbed as dark energy (DE). Despite the enormous
cosmological pieces of evidence, the origin as well as essential
characteristics of DE is still elusive and has become a source of
vivid debate. There are different DE models such as cosmological
constant $\Lambda$, quintessence energy, etc. that can effectively
describe the dynamics of the current universe. The cosmological
constant with negative pressure has the same value everywhere in
space, i.e., $\Lambda\approx1.3\times10^{-56}$cm$^{-2}$ \cite{25c}
which changes the spacetime structure of the compact objects. The
quintessence energy is inhomogeneous as well as dynamical scalar
field which can be characterized by the equation of state
$\omega_{q}= \frac{P_{q}}{\rho_{q}}$ with
$-1<\omega_{q}<-\frac{1}{3}$, where $P_{q}$ and $\rho_{q}$ denote
the pressure and energy density, respectively. Similar to the
cosmological constant, the cosmological horizon exists in the BH
spacetime immersed in the quintessential field.

In the presence of quintessential DE, the first-ever BH solution was
formulated by Kiselev \cite{25db}. He derived spherically symmetric
exact solutions of the field equations for charged as well as
uncharged BH surrounded by the quintessence. Following this
technique, various BH solutions have been constructed in the
background of quintessential field \cite{25dc}. Chen et al.
\cite{25da} examined the Hawking radiation spectra as well as
greybody factor for d-dimensional BH by using an equation of state
of quintessence matter and found that the luminosity of radiation
depends upon $|\omega_{q}|$. Hao et al. \cite{25daa} investigated
the absorption cross-section and absorption probability for the
Schwarzschild BH in the presence of quintessence matter. Saleh et
al. \cite{25ddaa} studied the quasinormal modes as well as Hawking
radiation for quintessential RN BH. Chakrabarty et al. \cite{25daaa}
analyzed the quasinormal modes as well as greybody factor for the
emission of scalar particles around a nonlinear magnetic-charged BH
surrounded by quintessence.

In this paper, we study an analytic form of the greybody factor in
the gravitational background corresponding to quintessential charged
rotating BH. The paper is outlined as follows. In the next section,
we evaluate the radial part of Klein-Gordon equation as well as
Schr$\ddot{o}$dinger equation to analyze the effective potential for
the massless scalar fields. Section \textbf{3} deals with two
analytic solutions of the radial equation of motion evaluated at two
specific radial regimes. In section \textbf{4}, we extrapolate these
asymptotic solutions to attain an analytic expression of the
greybody factor. We also calculate the energy emission rate and
absorption cross-section for the scalar field particles. Finally, we
summarize our results in the last section.

\section{Klein-Gordon Equation and Effective Potential}

The accelerated expansion of the universe could be the result of
quintessence matter which permeates the whole space. The
quintessential field around BH alters its spacetime properties as
well as asymptotic features of the cosmological horizon. Newman and
Janis \cite{25e} obtained Kerr BH solution from the Schwarzschild
spacetime by employing complex transformation within the framework
of Newman-Penrose formalism \cite{25f}. The same procedure was
adopted to generate the Kerr-Newman BH from the RN metric
\cite{25g}. The Newman-Janis algorithm (NJA) has been contemplated
as a favorable approach to obtain new rotating solutions of the
Einstein field equations \cite{25i, 25h}. Toshmatov et al. \cite{26}
derived the quintessential rotating BH solution by applying NJA on
the spherically symmetric BH. Using the same technique, Xu and Wang
\cite{26a} studied the Kerr-Newman solution in the presence of
quintessential DE. In the Boyer-Lindquist coordinates, the
Kerr-Neuman metric surrounded by quintessence can be expressed as
\begin{equation}\label{1}
ds^{2}=-F(r,\theta)dt^{2}+\frac{1}{G(r,\theta)}dr^{2}+\Sigma(r,\theta)
d\theta^2+H(r,\theta)d\phi^2-2K(r,\theta)dtd\phi,
\end{equation}
where
\begin{eqnarray}\nonumber
F(r,\theta)&=&\frac{\Delta(r)-a_{0}^{2}\sin^{2}\theta}{\Sigma(r,\theta)},
\quad G(r,\theta)=\frac{\Delta(r)}{\Sigma(r,\theta)}, \quad
\Sigma(r,\theta)=r^{2}+a_{0}^{2}\cos^{2}\theta,\\
\nonumber
H(r,\theta)&=&\frac{\sin^{2}\theta\left((r^{2}+a_{0}^{2})^{2}
-\Delta(r)a_{0}^{2}\sin^{2}\theta\right)}{\Sigma(r,\theta)},\\
\nonumber
K(r,\theta)&=&\frac{a_{0}\sin^{2}\theta(r^{2}+a_{0}^{2}-\Delta(r))}
{\Sigma(r,\theta)},\quad \Delta(r)=r^{2}+a_{0}^{2}+Q^{2}-2rM-\alpha
r^{1-3\omega_{q}}.
\end{eqnarray}
Here $a_{0}$ corresponds to the rotation parameter, $M$ and $Q$ are
the gravitational mass and total charge of BH, respectively,
$\omega_{q}$ is the dimensionless state parameter and $\alpha$ is
the quintessence parameter which determines the magnitude of
quintessence field around a BH, satisfying the inequality \cite{26a}
\begin{equation}
\alpha\leq\frac{2}{1-3\omega_{q}}8^{\omega_{q}}.
\end{equation}
This relation holds when the cosmological horizon determined by
quintessential DE exists. It is noted that charge does not affect
the range of $\alpha$, it remains the same for charge as well as
uncharged scenario. For $\alpha=0$, the line element (\ref{1})
reduces to Kerr-Newman BH which further leads to Kerr solution in
the absence of charge parameter. The horizons can be computed by the
constraint
\begin{equation}\label{1'}
\Delta(r)=0=r^{2}-2rM+a_{0}^{2}+Q^{2}-\alpha r^{1-3\omega_{q}}.
\end{equation}

The most appropriate method to study perturbations near a spacetime
generated by a BH is to allow probe fields to be perturbed by such
spacetime without reacting on it. If there is no effect of the field
on the spacetime, the perturbations of BH can be studied not only by
adding the perturbation terms, but also by introducing fields to the
spacetime \cite{c}. In general, for a scalar field, this leads to
find solutions for the Klein-Gordon equation with a well-defined
boundary condition. To analyze the emission of scalar fields $\Psi$
from a BH, we first derive the Klein-Gordon equation of a scalar
wave propagating in the gravitational background (\ref{1}). We
assume that massless particles are only minimally coupled to gravity
and do not involve in any other interaction. In this scenario, the
equation of motion for the curved spacetime is expressed as
\begin{equation}\label{2}
\nabla_{\mu}\nabla^{\mu}\Psi=\partial_{\mu}[\sqrt{-g}g^{\mu\nu}
\partial_{\nu}\Psi(t,r,\theta,\phi)]=0,
\end{equation}
which, through Eq.(\ref{1}), reduces to
\begin{eqnarray}\nonumber
&&\sqrt{-g}\left(\frac{-H}{K^2+FH}\right)\partial_{tt}\Psi
+(\sqrt{-g}G\partial_{r}\Psi)_{,r}+(\sqrt{-g}\frac{1}{\Sigma}
\partial_{\theta}\Psi)_{,\theta}\\ \label{3}
&&+\sqrt{-g}\left(\frac{F}{K^2+FH}\right)\partial_{\phi\phi}\Psi
+2\sqrt{-g}\left(\frac{-K}{K^2+FH}\right)\partial_{t}\partial_{\phi}\Psi=0.
\end{eqnarray}
Using the separation of variables ansatz
\begin{equation}\nonumber
\Psi(t,r,\theta,\phi)=\exp(-\iota wt)\exp(\iota
m\phi)R_{wlm}(r)Q^{m}_{l}(\theta,a_{0}w),
\end{equation}
where $w$ denotes the frequency of wave and
$Q^{m}_{l}(\theta,a_{0}w)$ corresponds to the angular spheroidal
functions \cite{24}, $R_{wlm}$ and $Q^{m}_{l}$ are obtained as the
solutions of the following decoupled equations
\begin{eqnarray}\nonumber
&\frac{\partial}{\partial r}(\Delta\frac{\partial R_{wlm}}{\partial
r})
+\left[\frac{1}{\Delta}(w^2(r^2+a_{0}^{2})^2+a_{0}^2m^{2}-2a_{0}wm(2rM-Q^{2}
+\alpha r^{1-3\omega_{q}}))\right.&\\\label{5} &\left.-a_{0}^2w^2
-\lambda_{l}^{m}\right]R_{wlm}=0,& \\\label{6}
&\frac{1}{\sin\theta}\frac{\partial}{\partial\theta}\left(\sin\theta
\frac{\partial
Q^{m}_{l}}{\partial\theta}\right)+(-\frac{m^{2}}{\sin^{2}\theta}
+w^{2}a_{0}^{2}\cos^{2}\theta+\lambda_{l}^{m})Q^{m}_{l}(\theta,a_{0}w)=0.&
\end{eqnarray}
Here $\lambda_{l}^{m}$ are the angular eigenvalues whose analytic
form in terms of power series \cite{22} can be written as
\begin{equation}\label{7}
\lambda_{l}^{m}=\sum_{n=0}^{\infty}\textit{f}_{~n}^{~lm}(a_{0}w)^{n}.
\end{equation}
The angular eigenvalue provides a connection between radial and
angular equations. In general, its analytic expression cannot be
written in a closed form. For simplicity, it is sufficient to keep
the finite number of terms and truncate the series at fourth order
given as follows
\begin{equation}\label{8}
\lambda_{l}^{m}=l(l+1)+\frac{2m^{2}-2l(l+1)+1}{(2l-1)(2l+3)}
(a_{0}w)^{2}+O((a_{0}w)^{4}),
\end{equation}
with $\textit{f}_{~1}^{~lm}=\textit{f}_{~3}^{~lm}=0$. The parameter
$l$ depicts the orbital angular momentum with non-negative integral
values and $m$ takes any integer value providing $l\geq |m|$ and
$\frac{l-|m|}{2}\in \{0,\mathbb{Z}^{+}\}$.

To derive the greybody factor for the massless scalar fields, we
determine an analytic solution of radial equation (\ref{5}) by using
the above-mentioned power series expression. Before attempting to
solve it analytically, we first analyze the profile of effective
potential which characterizes the emission process. Defining a new
radial function
\begin{equation}\label{9}
R_{wlm}(r)=\frac{U_{wlm}(r)}{\sqrt{r^2+a_{0}^{2}}},
\end{equation}
and use tortoise coordinate $x_{*}$ as
\begin{equation}\label{10}
\frac{dx_{*}}{dr}=\frac{r^{2}+a_{0}^2}{\Delta(r)},
\end{equation}
we get the following relations
\begin{equation}\label{11}
\frac{d}{dx_{*}}=\frac{\Delta(r)}{r^{2}+a_{0}^2}\frac{d}{dr}, \quad
\frac{d^2}{dx_{*}^{2}}=\left(\frac{\Delta(r)}{r^{2}+a_{0}^2}\right)^2
\frac{d^2}{dr^2}+\left(\frac{\Delta(r)}{r^{2}+a_{0}^2}\right)\frac{d}{dr}
\left(\frac{\Delta(r)}{r^{2}+a_{0}^2}\right)\frac{d}{dr}.
\end{equation}
The tortoise coordinate extends the range of the model between
$-\infty$ to $\infty$ whereas the Regge-Wheeler equation (\ref{5})
is confined only to regions located outside the BH horizon. In this
scenario, Eq.(\ref{5}) can be rewritten in the standard
Schr$\ddot{o}$dinger equation as
\begin{equation}\label{12}
(\frac{d^2}{dx_{*}^2}-V_{eff})U_{wlm}(r)=0,
\end{equation}
where the effective potential has the form
\begin{eqnarray}\nonumber
&V_{eff}=\{(r^2+a^2)^{\frac{1}{2}}\frac{d}{dr}\left[\frac{r\Delta(r)}
{(r^{2}+a_{0}^2)^{\frac{3}{2}}}\right]
-\frac{1}{\Delta}[w^2(r^2+a_{0}^{2})^2+a_{0}^2m^{2}-2a_{0}wm&\\
\nonumber &\times(2Mr-Q^{2}+\alpha r^{1-3\omega_{q}})]+
a_{0}^{2}w^2+l(l+1)+\frac{2m^{2}-2l(l+1)+1}{(2l-1)(2l+3)}
(a_{0}w)^{2}\}\frac{\Delta(r)}{(r^{2}+a_{0}^2)^2}.&\\\label{13}
\end{eqnarray}
\begin{figure}\center
\epsfig{file=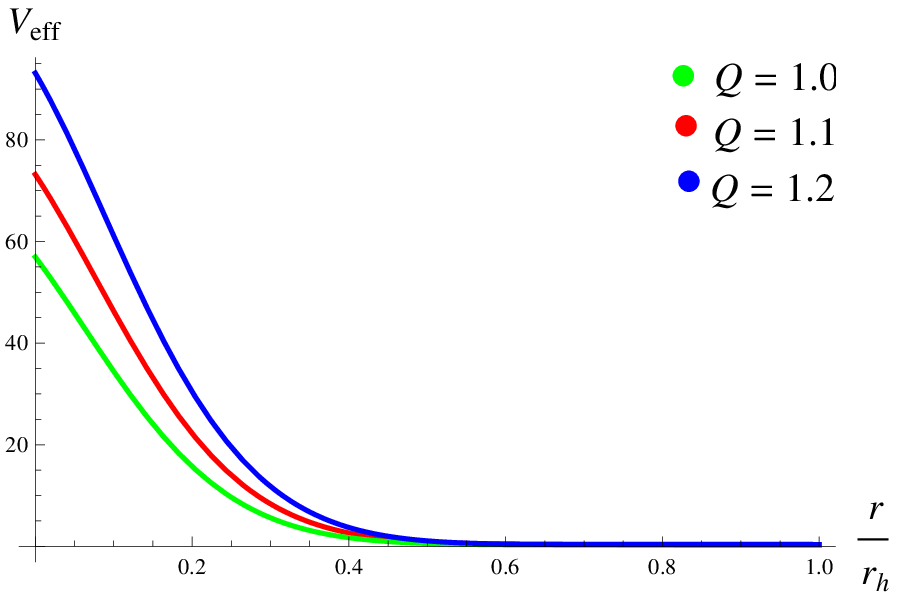,width=0.5\linewidth}\epsfig{file=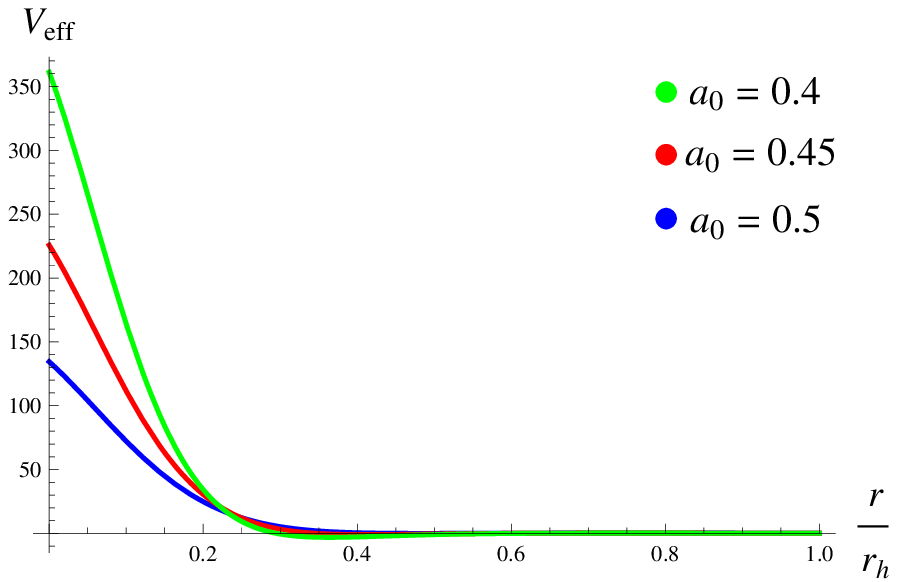,width=0.5\linewidth}\\
\caption{Effective potential for massless scalar fields
corresponding to $a_{0}=0.6$ (left plot) and  $Q=1$ (right plot)
with $m=l=M=1$, $\alpha=0.01$, $\omega_{q}=-0.6$ and $w=0.1$.}
\end{figure}
\begin{figure}\center
\epsfig{file=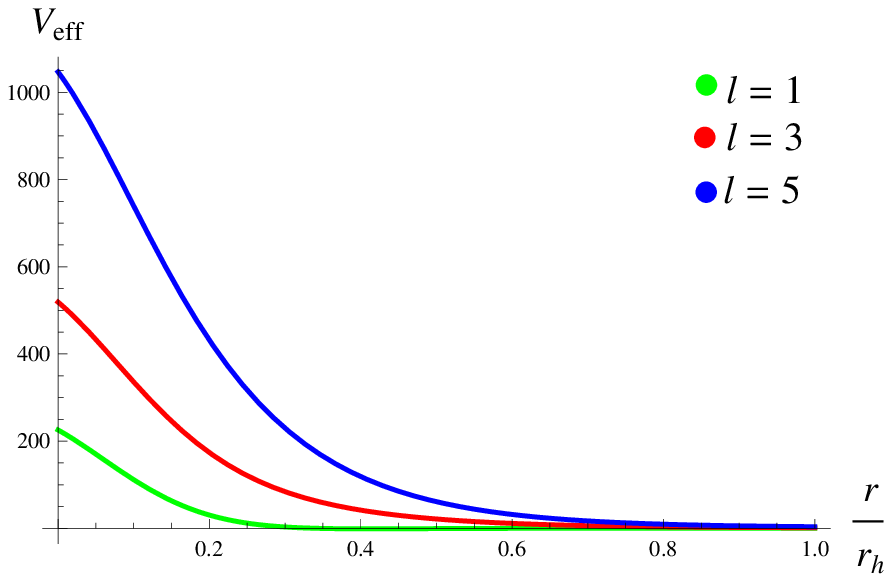,width=0.5\linewidth}\epsfig{file=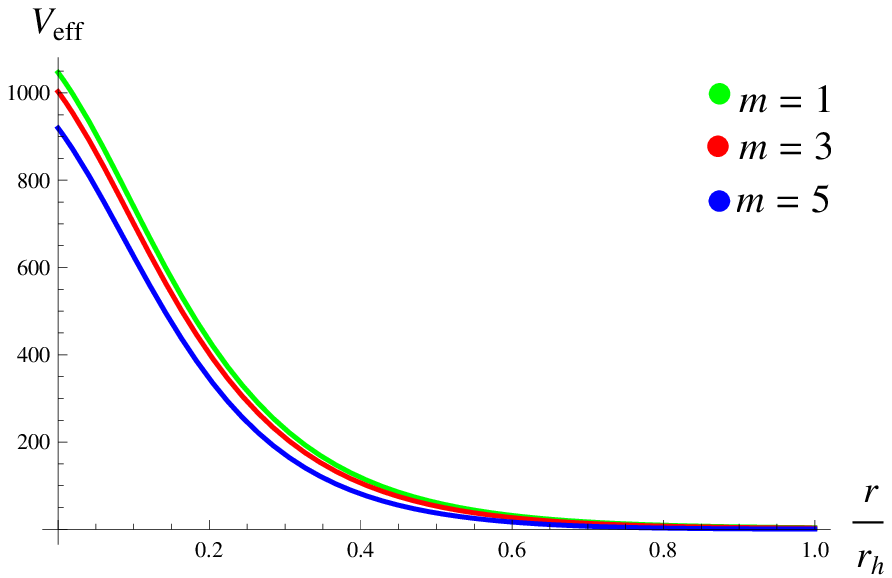,width=0.5\linewidth}\\
\caption{Effective potential for massless scalar fields
corresponding to $m=1$ (left plot) and  $l=5$ (right plot) with
$Q=M=1$ and $w=0.1$, $a_{0}=0.45$, $\omega_{q}=-0.6$ and
$\alpha=0.01$.}
\end{figure}

We see that the effective potential approaches to zero as
$x_{*}\rightarrow\pm\infty$. For graphical analysis, we display the
dependence of effective potential on different parameters in the
form of Figures \textbf{1-2}. Primarily, we set the parameters
$m=l=1=M$, $\alpha=0.01$, $\omega_{q}=-0.6$ and sketch the plots for
different choices of charge and rotation parameters. It is found
that the gravitational barrier increases gradually with the increase
in charge quantity (left plot of Figure \textbf{1}) which shows that
the inclusion of charge enhances the gravitational pull of BH. Thus,
both forces (electromagnetic and gravitational forces) act in the
same direction which ultimately increase the effective potential and
reduce the emission of scalar fields. As a result, the greybody
factor will decrease corresponding to larger choices of $Q$. In the
right plot of Figure \textbf{1}, we display the profile of effective
potential in terms of rotation parameter. We observe that the
potential barrier increases for smaller modes of $a_{0}$ leading to
the deduction of emission process. The dependence of the
gravitational barrier on angular momentum numbers is shown in Figure
\textbf{2}. It is noted that higher values of the potential are
obtained for larger modes of partial wave $l$ whereas the parameter
$m$ has an inverse impact on the effective potential, i.e., smaller
values of $m$ yield higher spikes of the potential.

\section{Greybody Factor}

In this section, we compute an analytic expression of the greybody
factor by solving the radial equation at specific radial regimes
such as close to the BH horizon and cosmological horizon determined
by the quintessence. We then use a semi-classical approach to
smoothly match these solutions in the low rotation regions.

\subsection{Analytic Solutions}

For the domain near to the BH event horizon $r\sim r_{h}$, we employ
the transformation
\begin{equation}\label{14}
r\rightarrow S=\frac{r^{2}+a_{0}^{2}+Q^{2}-2rM-\alpha
r^{1-3\omega_{q}}}{r^2+a_{0}^{2}-\alpha r^{1-3\omega_{q}}},
\end{equation}
which satisfies the relation
\begin{equation}\label{15}
\frac{dS}{dr}=\frac{(1-S)U(r_{h})}{r_{h}(r_{h}^{2}+a_{0}^2-\alpha
r_{h}^{1-3\omega_{q}})},
\end{equation}
with
\begin{equation}\label{16}
U(r_{h})=\frac{2Mr_{h}(r_{h}^2-a_{0}^{2}+3\alpha\omega_{q}r_{h}^{1-3\omega_{q}})
+Q^{2}r_{h}(-2r_{h}+\alpha(1-3\omega_{q})r_{h}^{-3\omega})}{2Mr_{h}-Q^2}.
\end{equation}
Using the above expressions in Eq.(\ref{5}), we have
\begin{eqnarray}\label{17'}
S(1-S)\frac{d^{2}R_{wlm}}{dS^{2}}+(1-C^{*}S)\frac{dR_{wlm}}{dS}
+\frac{1}{U^{2}(r_{h})(1-S)}[\frac{\chi^{*}}{S}-\lambda_{h}^{*}]R_{wlm}=0,
\end{eqnarray}
where
\begin{eqnarray}\label{18}
C^{*}&=&\frac{-2M(r_{h}^{2}+a_{0}^{2}-\alpha
r_{h}^{1-3\omega_{q}})}{2M(r_{h}^2-a_{0}^{2}+3\alpha\omega_{q}
r_{h}^{1-3\omega_{q}})+Q^{2}(-2r_{h}+\alpha(1-3\omega_{q})
r_{h}^{-3\omega})},\\ \label{19}
\chi^{*}&=&r_{h}^{2}[w^2(r_{h}^2+a_{0}^{2})^2
+a_{0}^{2}m^{2}-2a_{0}wm(2Mr_{h}-Q^{2}+\alpha
r_{h}^{1-3\omega_{q}})],
\end{eqnarray}
and
\begin{eqnarray}\label{20}
\lambda_{h}^{*}=r_{h}^{2}(r_{h}^{2}+a_{0}^2-\alpha
r_{h}^{1-3\omega_{q}})[a_{0}^{2}w^2+l(l+1)+\frac{2m^{2}-2l(l+1)+1}
{(2l-1)(2l+3)}(a_{0}w)^{2}].
\end{eqnarray}
Using the field redefinition
\begin{equation}\label{21}
R_{wlm}(S)=S^{\xi_{1}}(1-S)^{\eta_{1}}\hat{F}(S),
\end{equation}
Eq.(\ref{17'}) reduces to hypergeometric differential equation
\begin{eqnarray}\nonumber
&&S(1-S)\frac{d^{2}\hat{F}}{dS^{2}}+[1+2\xi_{1}-(2\xi_{1}+2\eta_{1}
+C^{*})S]\frac{d\hat{F}}{dS} +(\frac{\xi_{1}^2}{S}-\xi_{1}^2+\xi_{1}\\
\nonumber&&-2\xi_{1}\eta_{1}-\eta_{1}^2+\frac{\eta_{1}^2}{1-S}
-\frac{2\eta_{1}}{1-S}+\eta_{1}-\xi_{1}C^{*}+\frac{\eta_{1}C^{*}}
{1-S}-\eta_{1}C^{*}\\ \label{24}&&
+\frac{\chi^{*}}{U^{2}S}+\frac{\chi^{*}}{U^{2}(1-S)}
-\frac{\lambda_{h}^{*}}{U^{2}(1-S)})\hat{F}=0,
\end{eqnarray}
where
\begin{eqnarray}\label{25}
\hat{a}_{1}=\xi_{1}+\eta_{1}+C^{*}-1, \quad
\hat{b}_{1}=\xi_{1}+\eta_{1}, \quad \hat{c}_{1}=1+2\xi_{1}.
\end{eqnarray}

The power coefficients $\xi_{1}$ and $\eta_{1}$ can be computed by
algebraic equations, namely,
\begin{eqnarray}\label{26}
\xi_{1}^{2}+\frac{\chi^{*}}{U^{2}}&=&0,\\\label{27}
\eta_{1}^{2}+\eta_{1}(C^{*}-2)+\frac{\chi^{*}}{U^{2}}
-\frac{\lambda_{h}^{*}}{U^{2}}&=&0.
\end{eqnarray}
The radial equation of motion together with
Eqs.(\ref{25})-(\ref{27}) leads to
\begin{equation}\label{28}
S(1-S)\frac{d^{2}\hat{F}}{dS^2}+[\hat{c}_{1}-(1+\hat{a}_{1}
+\hat{b}_{1})S]\frac{d\hat{F}}{dS}-\hat{a}_{1}\hat{b}_{1}\hat{F}(S)=0.
\end{equation}
In terms of hypergeometric function, the general solution of
Eq.(\ref{24}) in near-horizon regime can be written as
\begin{eqnarray}\nonumber
(R_{wlm})_{NH}(S)&=&\tilde{A}_{1}S^{\xi_{1}}(1-S)^{\eta_{1}}\hat{F}
(\hat{a}_{1},\hat{b}_{1},\hat{c}_{1};S)+\tilde{A}_{2}S^{-\xi_{1}}\\
\label{29}
&\times&(1-S)^{\eta_{1}}\hat{F}(\hat{a}_{1}-\hat{c}_{1}+1,\hat{b}_{1}
-\hat{c}_{1}+1,2-\hat{c}_{1};S),
\end{eqnarray}
where $\tilde{A}_{1}$ and $\tilde{A}_{2}$ are arbitrary constants
with
\begin{eqnarray}\label{29'}
\xi_{1}^{\pm}&=&\pm\iota\frac{\sqrt{\chi^{*}}}{U(r_{h})},\\
\label{30}
\eta_{1}^{\pm}&=&\frac{1}{2}[(2-C^{*})\pm\sqrt{(2-C^{*})^2
-4(\frac{\chi^{*}}{U^2}-\frac{\lambda_{h}^{*}}{U^{2}})}].
\end{eqnarray}
Employing the boundary constraint that no outgoing waves are found
to be near the BH horizon, we can set either $\tilde{A}_{1}=0$ or
$\tilde{A}_{2}=0$ relying on the choice of coefficient $\xi_{1}$.
For both values of $\xi_{1}$, the constants become indistinguishable
from each other so that we choose $\xi_{1}=\xi_{1}^{-}$ and have
$\tilde{A}_{2}=0$. Similarly, the sign of $\eta_{1}$ can be decided
by the convergence property of hypergeometric function which demands
that we set $\eta_{1}=\eta_{1}^{-}$. The overall solution in the
near-horizon limit can be expressed as
\begin{equation}\label{30'}
(R_{wlm})_{NH}(S)=\tilde{A}_{1}S^{\xi_{1}}(1-S)^{\eta_{1}}\hat{F}
(\hat{a}_{1},\hat{b}_{1},\hat{c}_{1};S).
\end{equation}

Our next target is to solve the radial equation close to the
quintessence horizon $r_{q}$. Here, we will adopt the same procedure
as for BH horizon and replace the radial function $\Delta(r)$ with
$T(r)$ given by
\begin{equation}\label{31}
T(r)=1+\frac{a_{0}^2}{r^{2}}+\frac{Q^2}{r^{2}}-\alpha
r^{-1-3\omega_{q}},
\end{equation}
such that
\begin{equation}\label{32}
\frac{dT}{dr}=\frac{(1-T)D}{r},
\end{equation}
where
\begin{equation}\label{33}
D(r)=\frac{-2a_{0}^{2}-2Q^{2}+\alpha(1+3\omega_{q})r^{1-3\omega_{q}}}
{-a_{0}^2-Q^{2}+\alpha r^{1-3\omega_{q}}}.
\end{equation}
In the quintessential field, the radial equation of motion takes the
form
\begin{eqnarray}\label{34}
&T(1-T)\frac{d^{2}R_{wlm}}{dT^{2}}+(1-T)\frac{dR_{wlm}}{dT}
+[\frac{\chi_{q}^{*}}{D^{2}(1-T)T}-\frac{\lambda_{q}^{*}}{D^{2}(1-T)}]=0,&
\end{eqnarray}
with
\begin{eqnarray}\label{35}
\chi^{*}_{q}&=&r_{q}^{2}[w^2(r_{q}^2+a_{0}^{2})^2
+a_{0}^{2}m^{2}-2a_{0}wm(2Mr_{q}-Q^{2}+\alpha r_{q}^{1-3\omega_{q}})],\\
\label{36}
\lambda_{q}^{*}&=&r_{q}^{2}[a_{0}^{2}w^2+l(l+1)+\frac{2m^{2}-2l(l+1)+1}
{(2l-1)(2l+3)}(a_{0}w)^{2}].
\end{eqnarray}
We use the field redefinition
\begin{equation}\label{37}
R_{wlm}(T)=T^{\xi_{2}}(1-T)^{\eta_{2}}\hat{F}(T),
\end{equation}
which reduces Eq.(\ref{34}) to a hypergeometric equation with the
indices
\begin{eqnarray}\label{38}
\hat{a}_{2}=\xi_{2}+\eta_{2}=\hat{b}_{2}, \quad
\hat{c}_{2}=1+2\xi_{2}.
\end{eqnarray}
In this scenario, the power coefficients $\xi_{2}$ and $\eta_{2}$
can be computed as
\begin{eqnarray}\label{39}
\xi_{2}^{2}+\frac{\chi^{*}_{q}}{D^{2}}&=&0,\\\label{39'}
\eta_{2}^{2}+\frac{\chi^{*}_{q}}{D^{2}}
-\frac{\lambda_{q}^{*}}{D^{2}}&=&0.
\end{eqnarray}
For the quintessence cosmological horizon regime, the analytic
solution of Eq.(\ref{34}) in terms of hypergeometric function can be
written as
\begin{eqnarray}\nonumber
(R_{wlm})_{q}(T)&=&\hat{B}_{1}T^{\xi_{2}}(1-T)^{\eta_{2}}\hat{F}
(\hat{a}_{2},\hat{b}_{2},\hat{c}_{2};T)+\hat{B}_{2}T^{-\xi_{2}}\\
\label{40}
&\times&(1-T)^{\eta_{2}}\hat{F}(\hat{a}_{2}-\hat{c}_{2}+1,
\hat{b}_{2}-\hat{c}_{2}+1,2-\hat{c}_{2};T),
\end{eqnarray}
where $\hat{B}_{1}$ and $\hat{B}_{2}$ represent the arbitrary
constants. Here, we again opt negative values of $\xi_{2}$ and
$\eta_{2}$ to assure the convergence criterion of the hypergeometric
function.

\section{Matching to an Intermediate Regime}

In order to derive an analytical solution for the complete range of
$r$, we must ensure the smooth matching of two asymptotic solutions
$(R_{wlm})_{NH}$ and $(R_{wlm})_{q}$ at some intermediate region of
the radial coordinate. Starting form the near-horizon solution, we
first stretch the argument of hypergeometric function by replacing
$S$ with $1-S$ as
\begin{eqnarray}\nonumber
(R_{wlm})_{NH}(S)&=&\tilde{A}_{1}S^{\xi_{1}}(1-S)^{\eta_{1}}
\{\frac{\Gamma(\hat{c}_{1})\Gamma(\hat{c}_{1}-\hat{a}_{1}
-\hat{b}_{1})}{\Gamma(\hat{c}_{1}-\hat{a}_{1})\Gamma(\hat{c}_{1}
-\hat{b}_{1})}\hat{F}(\hat{a}_{1},\hat{b}_{1},\hat{c}_{1};1-S)\\
\nonumber &+&
(1-S)^{\hat{c}_{1}-\hat{a}_{1}-\hat{b}_{1}}\frac{\Gamma(\hat{c}_{1})
\Gamma(\hat{a}_{1}+\hat{b}_{1}-\hat{c}_{1})}{\Gamma(\hat{a}_{1})
\Gamma(\hat{b}_{1})}\\\label{41}&\times&\hat{F}(\hat{c}_{1}-\hat{a}_{1},
-\hat{b}_{1}+\hat{c}_{1},\hat{c}_{1}-\hat{b}_{1}-\hat{a}_{1}+1;1-S)\}.
\end{eqnarray}
Using Eq.(\ref{1'}), the function $S(r)$ can be rewritten as
\begin{equation}\label{42}
S(r)=1-\frac{2Mr-Q^2}{r^2+a_{0}^{2}-\alpha r^{1-\omega_{q}}}.
\end{equation}
For the limiting value $r>>r_{h}$ and $f\rightarrow 1$, the
stretched near-horizon solution has the form
\begin{equation}\label{43}
(1-S)^{\eta_{1}}\simeq\Big(\frac{r_{h}(1+a_{*}^{2}+Q_{*}^{2}-\alpha
r_{h}^{-1-3\omega_{q}})}{r}\Big)^{\eta_{1}}
\simeq\Big(\frac{r_{h}(1+a_{*}^{2}+Q_{*}^{2}-\alpha
r_{h}^{-1-3\omega_{q}})}{r}\Big)^{-l},
\end{equation}
and
\begin{eqnarray}\nonumber
(1-S)^{\eta_{1}+\hat{c}_{1}-\hat{a}_{1}-\hat{b}_{1}}
&\simeq&\left(\frac{r_{h}(1+a_{*}^{2}+Q_{*}^{2}-\alpha
r_{h}^{-1-3\omega_{q}})}{r}\right)^{-\eta_{1}+2-C^{*}}\\
\label{44}&\simeq&\left(\frac{r_{h}(1+a_{*}^{2}+Q_{*}^{2}-\alpha
r_{h}^{-1-3\omega_{q}})}{r}\right)^{l+1},
\end{eqnarray}
with $a_{*}= \frac{a_{0}}{r_{h}}$ and $Q_{*}= \frac{Q}{r_{h}}$. It
is worthwhile to mention here that all the mentioned-limitations are
valid for the smaller choices of charge and rotation parameter.
These approximations confine the validity of our results in the
low-energy region. Note that all the approximations are not done in
the argument of gamma function to enhance the efficiency of our
analytical solutions. The near-horizon solution, in an intermediate
region, takes the form
\begin{equation}\label{45}
(R_{wlm})_{NH}(r)=\hat{\mathcal{A}}_{1}\left(\frac{r}{r_{h}}\right)^{l}
+\hat{\mathcal{A}}_{2}\left(\frac{r}{r_{h}}\right)^{-(l+1)},
\end{equation}
with
\begin{eqnarray}\label{46}
\hat{\mathcal{A}}_{1}&=&\tilde{A}_{1}[(1+a_{*}^2+Q_{*}^{2}-\alpha
r_{h}^{-1-3\omega_{q}})]^{\eta_{1}}\frac{\Gamma(\hat{c}_{1})
\Gamma(\hat{c}_{1}-\hat{a}_{1}-\hat{b}_{1})}{\Gamma(\hat{c}_{1}
-\hat{a}_{1})\Gamma(\hat{c}_{1}-\hat{b}_{1})},\\\label{47}
\hat{\mathcal{A}}_{2}&=&\tilde{A}_{1}[(1+a_{*}^2+Q_{*}^{2}-\alpha
r_{h}^{-1-3\omega_{q}})]^{(-\eta_{1}-C^{*}+2)}\frac{\Gamma(\hat{c}_{1})
\Gamma(\hat{a}_{1}+\hat{b}_{1}-\hat{c}_{1})}{\Gamma(\hat{a}_{1})
\Gamma(\hat{b}_{1})}.
\end{eqnarray}

Next, we turn our attention towards the quintessence horizon
solution and shift the solution towards the smaller values of $r$ by
exchanging the argument of hypergeometric function from $T$ to
$1-T$. Setting $T(r_{q})\rightarrow 0$ leads to
\begin{eqnarray}\label{48}
(1-T)^{\eta_{2}}&\simeq&
\left(\frac{r}{r_{q}}\right)^{(-1-3\omega_{q})\eta_{2}} \simeq
\left(\frac{r}{r_{q}}\right)^{-(l+1)},\\ \label{49}
(1-T)^{\eta_{2}+\hat{c}_{2}-\hat{a}_{2}-\hat{b}_{2}}&\simeq&
\left(\frac{r}{r_{q}}\right)^{(-1-3\omega_{q})(\eta_{2}+\hat{c}_{2}
-\hat{a}_{2}-\hat{b}_{2})}\simeq \left(\frac{r}{r_{q}}\right)^{l},
\end{eqnarray}
which remain valid for smaller values of $a_{0}$ and $Q$. Under
these limitations, the solution of Eq.(\ref{40}) turns out to be
\begin{eqnarray}\label{50}
(R_{wlm})_{q}(r)&=&(\hat{\mathcal{H}}_{1}\hat{B}_{1}+\hat{\mathcal{H}}_{2}
\hat{B}_{2})\Big(\frac{r}{r_{q}}\Big)^{-(l+1)}+(\hat{\mathcal{H}}_{3}\hat{B}_{1}
+\hat{\mathcal{H}}_{4}\hat{B}_{2})\Big(\frac{r}{r_{q}}\Big)^{l},
\end{eqnarray}
where
\begin{eqnarray}\nonumber
\hat{\mathcal{H}}_{1}&=&\frac{\Gamma(\hat{c}_{2})\Gamma(\hat{c}_{2}-\hat{a}_{2}
-\hat{b}_{2})}{\Gamma(\hat{c}_{2}-\hat{a}_{2})\Gamma(\hat{c}_{2}
-\hat{b}_{2})}, \quad
\hat{\mathcal{H}}_{2}=\frac{\Gamma(2-\hat{c}_{2})\Gamma(\hat{c}_{2}-\hat{a}_{2}
-\hat{b}_{2})}{\Gamma(1-\hat{a}_{2})\Gamma(1-\hat{b}_{2})},\\
\nonumber
\hat{\mathcal{H}}_{3}&=&\frac{\Gamma(\hat{c}_{2})\Gamma(-\hat{c}_{2}+\hat{a}_{2}
+\hat{b}_{2})}{\Gamma(\hat{a}_{2})\Gamma(\hat{b}_{2})}, \quad
\hat{\mathcal{H}}_{4}=\frac{\Gamma(2-\hat{c}_{2})\Gamma(-\hat{c}_{2}+\hat{a}_{2}
-\hat{b}_{2})}{\Gamma(1-\hat{c}_{2}+\hat{a}_{2})\Gamma(1-\hat{c}_{2}
+\hat{b}_{2})}.
\end{eqnarray}
Now, we are in a position to evaluate the integration constants by
comparing the corresponding coefficients of two stretched solutions
(\ref{45}) and (\ref{50}) as both asymptotic solutions have the same
power coefficients, i.e., $r^{l}$ and $r^{-(l+1)}$. Thus, the
integration constants are found to be
\begin{eqnarray}\label{51}
\hat{B}_{1}&=&
\frac{\hat{\mathcal{A}}_{2}\hat{\mathcal{H}}_{4}-\hat{\mathcal{A}}_{1}
\hat{\mathcal{H}}_{2}}{\hat{\mathcal{H}}_{1}\hat{\mathcal{H}}_{4}
-\hat{\mathcal{H}}_{2}\hat{\mathcal{H}}_{3}},\quad\hat{B}_{2}=
\frac{\hat{\mathcal{A}}_{2}\hat{\mathcal{H}}_{3}-\hat{\mathcal{A}}_{1}
\hat{\mathcal{H}}_{1}}{\hat{\mathcal{H}}_{2}\hat{\mathcal{H}}_{3}
-\hat{\mathcal{H}}_{1}\hat{\mathcal{H}}_{4}}.
\end{eqnarray}
The expression of absorption probability for the emission of
massless scalar fields has the form
\begin{equation}\label{52}
|\tilde{\emph{A}}_{l,m}|^{2} =
1-\left|\frac{\hat{B}_{2}}{\hat{B}_{1}}\right|^{2},
\end{equation}
which, through Eq.(\ref{51}), gives rise to
\begin{eqnarray}\label{53}
|\tilde{\emph{A}}_{l,m}|^{2} &=&
1-\left|\frac{\hat{\mathcal{A}}_{2}\hat{\mathcal{H}}_{3}
-\hat{\mathcal{A}}_{1}\hat{\mathcal{H}}_{1}}
{\hat{\mathcal{A}}_{2}\hat{\mathcal{H}}_{4}
-\hat{\mathcal{A}}_{1}\hat{\mathcal{H}}_{2}}\right|^{2}.
\end{eqnarray}
The above relation specifying the emission of scalar fields from a
charged rotating BH surrounded by quintessence matter, remains valid
in low-charge and low-angular momentum regions.

Any traveling wave incoming towards a BH faces the effective
potential as a barrier which partially transmits it or partially
reflects it back. It is the relative relation between the effective
potential and frequency which decides either to reflect the wave or
move forward. For the region near to the event horizon defined by
$V_{eff}(r)<<w^{2}$, the wave may cross the barrier and will not be
reflected. In this scenario, the transmission coefficient will
approach to unity and the reflection parameter will almost equal to
zero. In the reverse case, when the height of potential is larger as
compared to the frequency, most of the part will be reflected and
some of its portion may cross the barrier through the tunneling
effect. In this case, the greybody factor shows a negative trend.
\begin{figure}\center
\epsfig{file=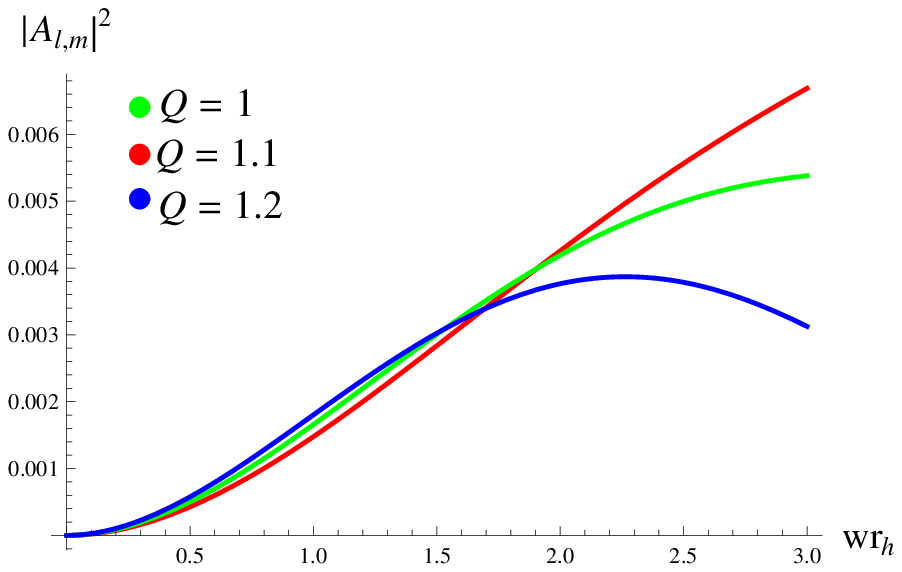,width=0.5\linewidth}\epsfig{file=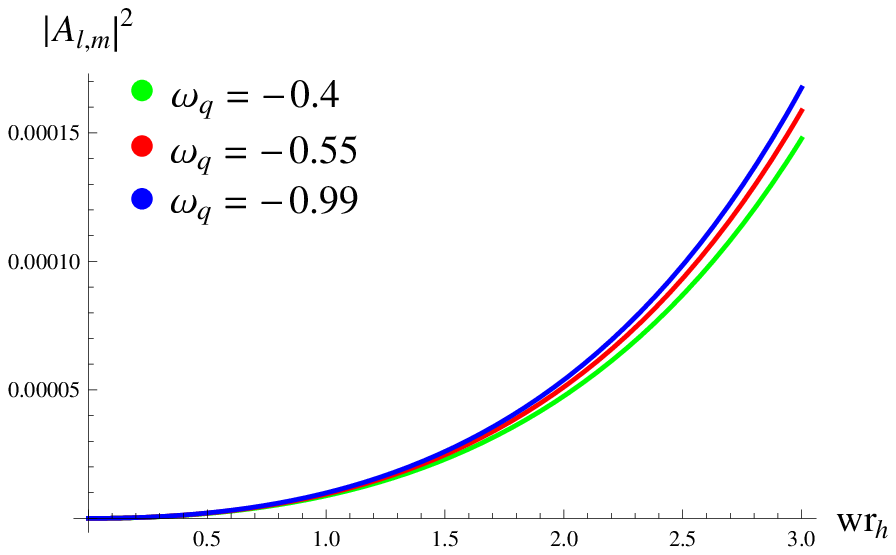,width=0.5\linewidth}\\
\caption{Greybody factor for massless scalar fields corresponding to
$\omega_{q}=-0.4$, (left plot) and $Q=0.01$ (right plot) with
$l=m=0$, $M=1$ and $a_{0}=0.01=\alpha$.}
\end{figure}

In order to analyze the significant features of greybody factor, we
sketch the expression (\ref{53}) versus dimensionless parameter
$wr_{h}$ and examine its dependence on topological parameters $(Q,
a_{0}, \alpha, \omega_{q})$ and angular momentum numbers $(l, m)$.
It is observed that the absorption probability interpolates smoothly
between $0$ and asymptotic value $1$. In Figure \textbf{3}, we plot
the graphs for different choices of charge and state parameters by
considering the other variables as fixed quantities. It is found
that the greybody factor gets suppressed for larger values of charge
(left plot) as expected from the effective potential plot (Figure
\textbf{1}) whereas the higher modes of state parameter depict a
slight increase in the absorption probability (right plot).
\begin{figure}\center
\epsfig{file=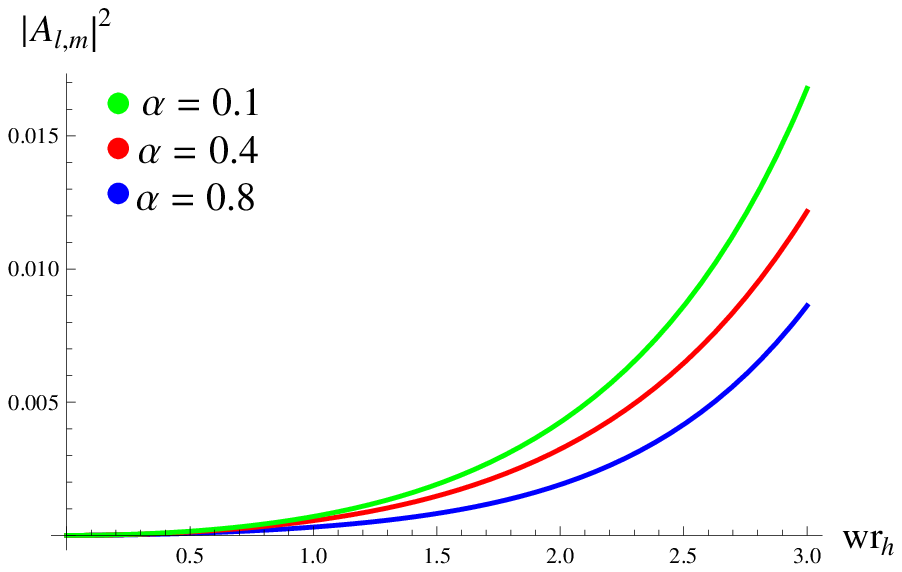,width=0.5\linewidth}\epsfig{file=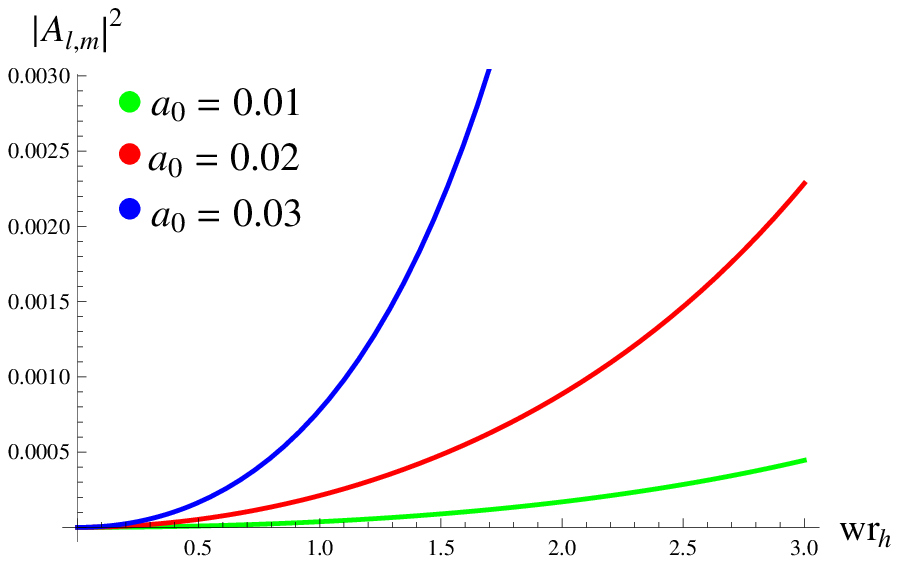,width=0.5\linewidth}\\
\caption{Greybody factor for massless scalar fields corresponding to
$a_{0}=0.03$, (left plot) and $\alpha=0.01$ (right plot) with
$l=m=0$, $M=1$, $Q=0.1$ and $\omega_{q}=-0.6$.}
\end{figure}

Figure \textbf{4} (left plot) indicates that increase in the
strength of quintessence matter $\alpha$ causes a reduction in the
emission of scalar fields. Moreover, we obtain that the greybody
factor increases comprehensively for the larger modes of rotation
parameter (right plot). The impact of positive as well as negative
values of $m$ on the greybody factor is displayed in Figure
\textbf{5}. It is noted that an increase in $m$ reduces the emission
process while the negative modes of $m$ yield higher values of
absorption probability. Finally, the effect of orbital angular
momentum on absorption probability is shown in Figure \textbf{6}.
This indicates that low partial wave ($l=1$) leads to smaller values
of the greybody factor while higher values of $l$ dominate in the
high-energy regions.
\begin{figure}\center
\epsfig{file=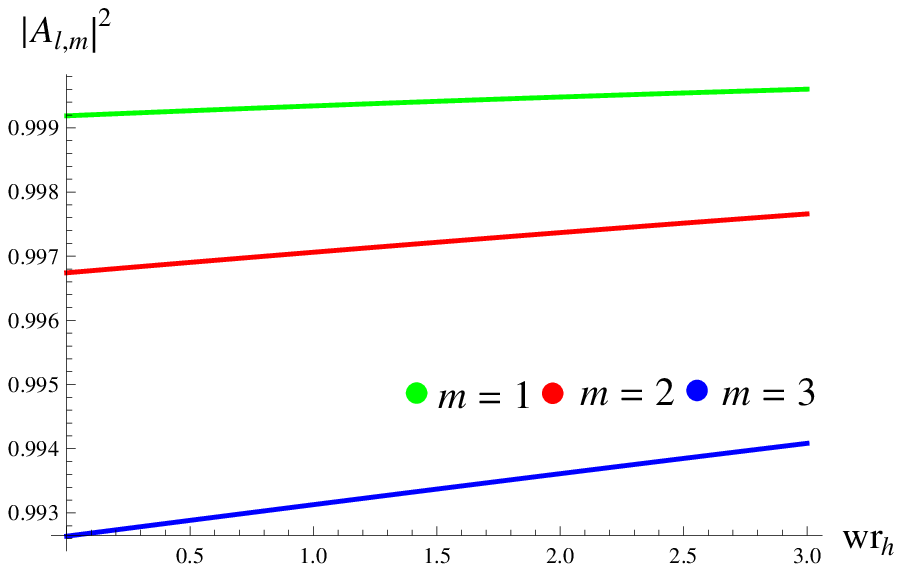,width=0.5\linewidth}\epsfig{file=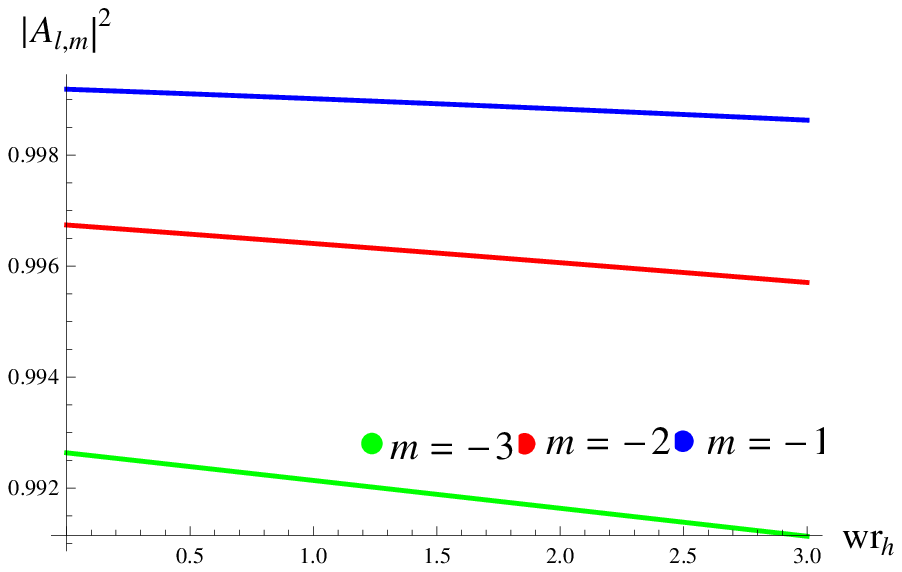,width=0.5\linewidth}\\
\caption{Greybody factor for massless scalar fields corresponding to
$l=3$, $\alpha=a_{0}=0.1=Q$, $M=1$ and $\omega_{q}=-0.4$ }
\end{figure}
\begin{figure}\center
\epsfig{file=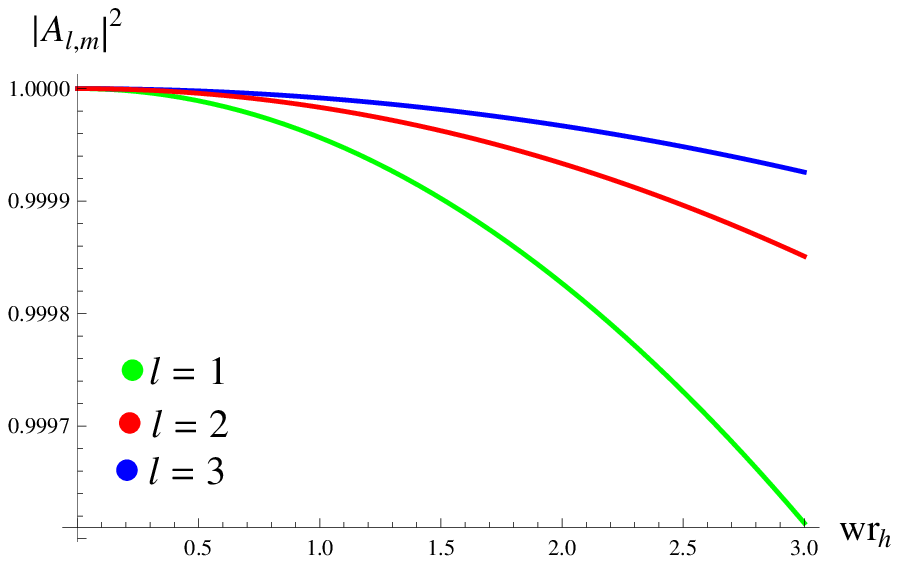,width=0.5\linewidth}\\
\caption{Greybody factor for massless scalar fields corresponding to
$m=0$, $\alpha=a_{0}=0.1=Q$, $M=1$ and $\omega_{q}=-0.4$ }
\end{figure}

The total amount of massless particles emitted from a BH per unit
time and frequency is given by
\begin{eqnarray}\label{54}
\frac{d^2\tilde{N}}{dtdw}=\frac{1}{2\pi}\sum_{l,m}\frac{1}
{e^{\frac{k}{T_{H}}-1}}|\tilde{\emph{A}}_{l,m}|^{2}, \quad
k=w-\frac{ma_{0}}{r_{h}^2+a_{0}^2}.
\end{eqnarray}
Moreover, the energy emission rate can be expressed as
\begin{eqnarray}\label{55}
\frac{d^2\tilde{E}}{dtdw}=\frac{1}{2\pi}\sum_{l,m}\frac{w}
{e^{\frac{k}{T_{H}}-1}}|\tilde{\emph{A}}_{l,m}|^{2},
\end{eqnarray}
which, through Eq.(\ref{53}), has the form
\begin{eqnarray}\label{56}
\frac{d^2\tilde{E}}{dtdw}=\frac{1}{2\pi}\sum_{l,m}\frac{w}
{e^{\frac{k}{T_{H}}-1}}\left(1-\left|\frac{\hat{\mathcal{A}}_{2}
\hat{\mathcal{H}}_{3}-\hat{\mathcal{A}}_{1}\hat{\mathcal{H}}_{1}}
{\hat{\mathcal{A}}_{2}\hat{\mathcal{H}}_{4}-\hat{\mathcal{A}}_{1}
\hat{\mathcal{H}}_{2}}\right|^{2}\right).
\end{eqnarray}
The dependence of greybody factor on particle as well as spacetime
properties changes the various emission rates, accordingly. The
absorption cross-section for charged rotating BH surrounded by
quintessential field is given as
\begin{eqnarray}\label{57}
\sigma=\frac{\pi}{w^2}\sum_{l,m}|\tilde{\emph{A}}_{l,m}|^{2}.
\end{eqnarray}
Using Eq.(\ref{53}), we have
\begin{eqnarray}\label{58}
\sigma=\frac{\pi}{w^2}\sum_{l,m}\left(1-\left|\frac{\hat{\mathcal{A}}_{2}
\hat{\mathcal{H}}_{3}-\hat{\mathcal{A}}_{1}\hat{\mathcal{H}}_{1}}
{\hat{\mathcal{A}}_{2}\hat{\mathcal{H}}_{4}
-\hat{\mathcal{A}}_{1}\hat{\mathcal{H}}_{2}}\right|^{2}\right).
\end{eqnarray}
The absorption cross-section as a function of incident frequency is
used to quantify the probability of a certain particle-particle
interaction such as scattering, electromagnetic absorption, etc. It
exhibits oscillations around the limit of geometrical optics which
is a characteristic of diffraction patterns \cite{45a}.

\section{Conclusions}

To examine the Hawking radiation spectra emitted from various BH
geometries, the greybody factors for different scalar fields have
intensively been studied. This paper formulates an analytic
expression of the greybody factor for charged rotating BH surrounded
by the quintessence, valid in low-energy approximation. Initially,
we have investigated the profile of effective potential which
originates the absorption probability. The radial equation of motion
has been solved analytically at two specific horizons to obtain
asymptotic solutions in the form of hypergeometric functions. We
have extrapolated these solutions and matched them smoothly to an
intermediate regime to obtain a general form of the greybody factor.
The energy emission rate and absorption cross-section have been
computed for the massless scalar fields.

It is found that the height of effective potential increases with
the gradual increase in charge. It is worthwhile to mention here
that the electromagnetic force enhances the gravitational pull of
the BH which ultimately minimizes the emission rate of Hawking
radiation (Figure \textbf{1}). The larger values of $a_{0}$ decrease
the gravitational barrier for the massless scalar fields whereas the
higher modes of $l$ have an inverse effect on the effective
potential.

The graphical analysis of absorption probability indicates its
positive range throughout the considered domain. An increase in $Q$
leads to the reduction of absorption probability (Figure \textbf{3})
which is also consistent with the literature \cite{45}. It is found
that the greybody factor gets suppressed for larger values of
$\alpha$ in comparison with \cite{25daaa} whereas the higher modes
of rotation parameter show a substantial increase in the emission
rate of massless scalar particles. For the orbital angular momentum,
partial wave with smaller values reduce the greybody factor as for
the rotating BH \cite{40}. It is observed that by taking $\alpha=0$,
$Q=0$ and $a_{0}=0$, the line-element reduces to the Kerr-Newman,
Kerr and Schwarzschild BH, respectively. Consequently, the
analytical expressions of the effective potential and greybody
factor reduce to the corresponding BH solutions which are in
well-agrement with the literature \cite{a}-\cite{c}. We conclude
that the inclusion of charge parameter in the presence of
quintessential field significantly affects the potential barrier as
well as greybody factor. It would be interesting to study the
quasinormal modes resonant phenomena in the background of
quintessential BH as done for other BH \cite{d}.

\vspace{0.5cm}

\textbf{Acknowledgement}

\vspace{0.5cm}

One of us (QM) would like to thank the Higher Education Commission,
Islamabad, Pakistan for its financial support through the
\emph{Indigenous Ph.D. Fellowship, Phase-II, Batch-III}.

\end{document}